\documentclass[aps,prc,twocolumn, superscriptaddress,showkeys,nofootinbib]{revtex4-1}  

\usepackage[utf8]{inputenc}

\usepackage{graphicx,color}
\usepackage{amsmath,amssymb,amsfonts}
\usepackage{hyperref}

\usepackage{xspace}	

\newcommand{\roots} {\mbox{$\sqrt{\textit{s}_{NN}}$}\xspace}

\def  \vn         {\mbox{$\textit{v}_{n}$}\xspace}

\def \cnm2   {\mbox{$\mathrm{C}_{nm}\lbrace 2 \rbrace$ }\xspace}
\def \cnm4   {\mbox{$\mathrm{C}_{nm}\lbrace 4 \rbrace$ }\xspace}

\def \sc23  {\mbox{$\mathrm{SC}(2,3)$ }\xspace}
\def \sc24  {\mbox{$\mathrm{SC}(2,4)$ }\xspace}

\def \nsc23 {\mbox{$\mathrm{NSC}(2,3)$}\xspace}
\def \nsc24 {\mbox{$\mathrm{NSC}(2,4)$}\xspace}

\def  \RuRu       {\mbox{$\rm Ru+Ru$ }\xspace}
\def  \ZrZr       {\mbox{$\rm Zr+Zr$ }\xspace}

\def  \RuRuA       {\mbox{$^{96}Ru+^{96}Ru$ }\xspace}
\def  \ZrZrA       {\mbox{$^{96}Zr+^{96}Zr$ }\xspace}

\begin{document}

\title{
Characterizing the initial and final state effects in isobaric collisions at energies available at the BNL Relativistic Heavy Ion Collider
}
\medskip
\author{Niseem~Magdy} 
\email{niseemm@gmail.com}
\affiliation{Center for Frontiers in Nuclear Science at the State University of New York, Stony Brook, NY 11794, USA}
\affiliation{Department of Chemistry, the State University of New York, Stony Brook, New York 11794, USA}

\begin{abstract}
%
%

The Multi-Phase Transport model (AMPT) is employed to predict Symmetric Correlations (SC), Asymmetric Correlations (ASC), Normalized Symmetric Correlations (NSC), and Normalized Asymmetric Correlations (NASC) in \RuRuA and \ZrZrA collisions at 200~GeV. Our study offers insights into the behavior of SC, ASC, NSC, and NASC, considering various nuclear structure scenarios to account for differences between the two isobars. Additionally, we emphasize the importance of detailed experimental measurements as they will serve as a critical constraint for refining the predictions of theoretical models. 
\end{abstract}
\maketitle
\section{Introduction}
The strongly coupled Quark Gluon Plasma(QGP) is expected to be created in the nuclear collisions at the Relativistic Heavy Ion Collider (RHIC) and the Large Hadron Collider (LHC). A comprehensive description and understanding of the QGP transport properties (i.e., the specific viscosity $\eta/s$) is essential to the ongoing high-energy nuclear research. Nevertheless, such a description demands a detailed understanding of the Heavy Ion Collisions (HIC) initial state ~\cite{Danielewicz:1998vz, Ackermann:2000tr, Adcox:2002ms, Heinz:2001xi, Hirano:2005xf, Huovinen:2001cy, Hirano:2002ds, Romatschke:2007mq, Luzum:2011mm, Song:2010mg, Qian:2016fpi, Schenke:2011tv, Teaney:2012ke, Gardim:2012yp, Lacey:2013eia, Magdy:2022ize}.
Many studies have highlighted the importance of anisotropic flow measurements to serve as a valuable input for examining both the initial conditions and transport properties of the QGP~\cite{Teaney:2003kp, Lacey:2006pn, Song:2011qa, Alver:2010gr, Schenke:2010rr, Shen:2010uy, Niemi:2012ry, Qin:2010pf, Magdy:2023owx}.

Indeed, considerable research efforts have provided a rich collection of knowledge about the HIC initial conditions and the QGP properties. This valuable knowledge has been obtained by conducting extensive investigations into the n$^{th}$ order flow harmonics.
{\color{black}Note that, the n$^{th}$ complex flow vector with $\Psi_{n}$ direction and $v_n$ magnitude is given as~\cite{Voloshin:1994mz,Poskanzer:1998yz}:
\begin{eqnarray}\label{Eq:Vn}
\textit{V}_{n} = \vn e^{i \textit{n} \Psi_{n}},
\end{eqnarray}
where $v_1$, $v_2$, and $v_3$ are the dipolar, elliptic, and triangular flow harmonics, respectively. The $\textit{V}_{n}$ gives the hydrodynamic response of the created medium to the n$^{th}$-order initial-state eccentricity $\varepsilon_{n}$~\cite{Gardim:2011xv, Niemi:2012aj}.}
These studies have explored the flow harmonics magnitudes, statistical variations, fluctuations and the correlations among them~\cite{STAR:2018fpo, ALICE:2016kpq, Adamczyk:2017hdl, Qiu:2011iv, Adare:2011tg, Aad:2014fla, Aad:2015lwa, Magdy:2018itt, Alver:2008zza, Alver:2010rt, Ollitrault:2009ie, Magdy:2022jai, STAR:2022gki, Adamczyk:2016gfs, STAR:2015rxv, Magdy:2018itt, Adamczyk:2015obl, Adamczyk:2016exq, Adam:2019woz, Song:2010mg, Niemi:2012aj, Gardim:2014tya, Fu:2015wba, Holopainen:2010gz, Qin:2010pf, Qiu:2011iv, Gale:2012rq, Liu:2018hjh, Magdy:2022ize}.%
In HIC, the structure of the colliding nuclei can impact the n$^{th}$ order flow harmonics $v_{n}$, which reflect the initial state eccentricities $\varepsilon_{n}$ dependence on the colliding nuclei structure~\cite{Jia:2021tzt, Giacalone:2021udy, Magdy:2022cvt, Magdy:2023fsp}. Such structure can be described with the woods-Saxon distribution for the nuclear density as:
\begin{eqnarray}\label{Eq:Rthita}
\rho(r, \theta, \phi) &\propto& \frac{1}{1+e^{(r-R(\theta, \phi))/a_0}}, \\
R(\theta, \phi) &=& R_0(1+\beta_2Y^0_2( \theta, \phi)+\beta_3Y^0_3( \theta, \phi)),
\end{eqnarray}
where $R(\theta, \phi)$ is the nuclear surface incorporating the quadrupole and octupole deformations controlled by $\beta_2$ and $\beta_3$, respectively, the parameters $R_0$ and $a_0$ describe the half-width radius and nuclear skin thickness~\cite{Filip:2007tj, Jia:2021wbq}.

Recent theoretical and experimental studies have suggested that investigating the ratios of the n$^{th}$-order flow harmonics (i.e., $\frac{v^{\rm Ru+Ru}_n}{v^{\rm Zr+Zr}_n}$) will provide essential insights on the structure of the colliding nuclei~\cite{Li:2018oec, Li:2019kkh, Xu:2021vpn, Xu:2021qjw, Xu:2021uar, Zhao:2022uhl, Xu:2022ikx, Wang:2023yis, Dimri:2023wup, Liu:2022xlm, Nie:2022gbg, Jia:2022qgl, Jia:2022qrq, Zhang:2022fou, Jia:2022iji, Liu:2022kvz, Magdy:2023fsp, Cheng:2023ucp, Zhao:2022mce}.
In the case of \RuRu and \ZrZr collisions, the hope was to impose constraints on the differences in the nuclear deformation and nuclear skin between them. Building these ratios stems from the expectation that the viscosity attenuation effects in these isobaric collisions should be similar~\cite {STAR:2019zaf}. In contrast, slight variations in their deformation and nuclear skin can lead to differences in the initial-state eccentricities~\cite{Jia:2021tzt, Giacalone:2021udy}, causing apparent disparities in the magnitudes of $v_2$ and $v_3$ between \RuRu and \ZrZr. 
Many of these investigations focused on the isobaric ratios only. In this work, we highlight the value of the observables sensitive to initial and final state effects.  

Reaching a comprehensive understanding of the QGP requires a simultaneous constraint on estimating the initial state eccentricities and their fluctuations and correlations as well as the final state viscous attenuation ($\eta/s$)~\cite{Schenke:2019ruo, Alba:2017hhe}. In a prior investigation~\cite{Magdy:2022ize}, we showed that such constraints can be achieved via data model comparisons of Symmetric Correlations (SC), Asymmetric Correlations (ASC)~\cite{Mordasini:2019hut, Moravcova:2020wnf, Li:2021nas, Bilandzic:2021rgb, Nijs:2021kvn, Zhang:2021kxj}, Normalized Symmetric Correlations (NSC), and Normalized Asymmetric Correlations (NASC)~\cite{Bilandzic:2021rgb, Bilandzic:2020csw, Taghavi:2020gcy, ALICE:2017fcd}. Consequently, in this work, we will extend the previous study of SC, ASC, NSC, and NASC~\cite{Taghavi:2020gcy, Bilandzic:2021rgb, Bilandzic:2020csw, Mordasini:2019hut, Li:2021nas, Moravcova:2020wnf} to the \RuRu and \ZrZr at  RHIC top energy. In addition, we systematically investigate the influence of the initial state deformation and nuclear skin using the AMPT model~\cite{Lin:2004en} on the SC, ASC, NSC, and NASC.

The motivation behind these investigations is to provide a systematic study of a comprehensive set of flow correlators that have the potential to constrain the HIC's initial and final effects simultaneously. In this work, we limit the presentation of the results to the system-size comparisons to provide the AMPT predictions of those flow observables. The paper is organized as follows. Section~\ref{sec:2} summarizes the theoretical model used to investigate the SC, ASC, NSC, and NASC and the details of the analysis method employed. The results are presented in Sec.~\ref{sec:3} followed by a summary and outlook in Sec.~\ref {sec:4}.

\section{Methodology} \label{sec:2}

\subsection{The AMPT Model}
In this investigation, we employ the multiphase transport model AMPT to study the SC, ASC, NSC, and NASC in $^{96}$Ru+$^{96}$Ru and $^{96}$Zr+$^{96}$Zr collisions at \roots = 200~GeV. 
The AMPT model~\cite{Lin:2004en} is a comprehensive simulation framework widely utilized to delve into the complexities of relativistic heavy-ion collisions~\cite{Lin:2004en, Ma:2016fve, Haque:2019vgi, Bhaduri:2010wi, Nasim:2010hw, Xu:2010du, Magdy:2020bhd, Guo:2019joy}. In this investigation, we generate AMPT events incorporating the string-melting option.
Within the AMPT framework, the Glauber model defines the shape and radial distribution of the colliding nuclei, employing a deformed Woods-Saxon distribution~\cite{Hagino:2006fj} as given in Eq.~\ref{Eq:Rthita}.
In AMPT simulations, the projectile and target nuclei are rotated event-by-event randomly along the polar and azimuthal directions. 
These analyses consider the intrinsic deformations and nuclear skin differences between $^{96}$Ru and $^{96}$Zr. Therefore, the Woods-Saxon parameter sets for $^{96}$Ru and $^{96}$Zr employed in our analysis are summarized in  Table~\ref{tab:1}~\cite{Moller:1993ed}.
{\Large
\begin{table}[h!]
\begin{center}
 \begin{tabular}{|c|c|c|c|c|}
 \hline 
  AMPT-set      &   $R_{0}$ &  $a_{0}$  & $\beta_{2}$  &  $\beta_{3}$   \\
  \hline
  $^{96}$Ru+$^{96}$Ru (Case-1)    &    5.09   &  0.46   &  0.162       &   0.00           \\
  \hline
  $^{96}$Zr+$^{96}$Zr (Case-2)    &    5.02   &  0.52   &  0.06        &   0.20           \\
 \hline
  $^{96}$Zr+$^{96}$Zr (Case-3)    &    5.09   &  0.46   &  0.06        &   0.20           \\
 \hline
  $^{96}$Zr+$^{96}$Zr (Case-4)    &    5.09   &  0.46   &  0.06        &   0.00           \\
 \hline
\end{tabular} 
\caption{ Summary of the Woods-Saxon parameters for $^{96}$Ru and $^{96}$Zr employed in the AMPT studies.}
\label{tab:1}
\end{center}
\end{table}
}

Within the AMPT framework, the HIJING model generates hadrons, which are then transformed into their constituent quarks and anti-quarks. The subsequent space-time evolution of these particles is determined using the ZPC Parton cascade model~\cite{Zhang:1997ej}, which incorporates the parton-scattering cross-section:
\begin{eqnarray} \label{eq:21}
\sigma_{pp} &=& \dfrac{9 \pi \alpha^{2}_{s}}{2 \mu^{2}},
\end{eqnarray}
where $\alpha_{s}$ denotes the QCD coupling constant and $\mu$ is the screening mass in the partonic matter; these parameters collectively shape the expansion dynamics of the collision systems~\cite{Zhang:1997ej}.  Collisions were  simulated for a fixed $\sigma_{pp}$ = 2.8~\cite{Xu:2011fi, Nasim:2016rfv}. Thus, the AMPT model amalgamates several contributing factors: (i) the initial-state eccentricity, (ii) the initial Parton-production stage governed by the HIJING model~\cite{Wang:1991hta, Gyulassy:1994ew}, (iii) a Parton-scattering stage, (iv) the process of hadronization through coalescence, and finally, (v) a phase that encompasses interactions among the generated hadrons~\cite{Li:1995pra}.\\

\subsection{Analyses Method}
The multi-particle correlation technique with traditional and two-subevents cumulants methods~\cite{Bilandzic:2010jr, Bilandzic:2013kga, Gajdosova:2017fsc, Jia:2017hbm} are used in this work. In this study, we used the two subevents with a pseudorapidity gap $\Delta\eta~=~\eta_{a}-\eta_{b}~ > 0.7$ (i.e., $\eta_{a}~ > 0.35$ and $\eta_{a}~ < -0.35$) for the two-, three-, and four-particle correlations. However, we used the traditional cumulant method to calculate the five- and six-particle correlations. More details on the methods used can be found in Ref~\cite{Magdy:2022ize}.

%
\begin{widetext}
\subsubsection{The k-particle symmetric correlations:}
The two-, four-, and six-particle symmetric correlations can be given using the multi-particle correlation techniques:
{ 
\begin{flalign}\label{eq:2-1}
    SC(n_1, -n_1)|_{2-Sub} ={}& \langle\langle  \cos(n_{1}\phi^{a}_{x1} - n_{1}\phi^{b}_{x2} \rangle\rangle, & \\  \nonumber
                ={}& \langle\langle  v_{n_{1}}^{2} \cos(n_{1}\psi_{n_{1}} - n_{1}\psi_{n_{1}} )\rangle\rangle, &\\  \nonumber
                ={}& \langle\langle  v^{2}_{n_{1}} \rangle\rangle, &
\end{flalign}
}
{\color{black}where, $n_{i}$ is the harmonic order, $\phi^{a}_{x1}$ is the azimuthal angle of the particle $x_1$ in the subevent $a$, and the $\langle\langle  \rangle\rangle$ defines the averaging first over particles in an event and then over events.}
{ 
\begin{flalign}\label{eq:2-3}
    SC(n_1, n_2, -n_1, -n_2)_{2-Sub} ={}& \langle\langle  \cos(n_{1}\phi^{a}_{x1} + n_{2}\phi^{a}_{x2} - n_{1}\phi^{b}_{x3} - n_{2}\phi^{b}_{x4} \rangle\rangle, & \\  \nonumber
                       ={}& \langle\langle  v_{n_{1}}^{2} v_{n_{2}}^{2} \cos(n_{1}\psi_{n_{1}} + n_{2}\psi_{n_{2}} - n_{1}\psi_{n_{1}} -  n_{2}\psi_{n_{2}} )\rangle\rangle, &\\  \nonumber
                       ={}& \langle\langle  v^{2}_{n_{1}}  v^{2}_{n_{2}} \rangle\rangle, &
\end{flalign}
}
{
\begin{flalign}\label{eq:2-6}
   SC(n_1, n_2, n_3, -n_1, -n_2, -n_3)_{Trad} ={}& \langle\langle  \cos(n_{1}\phi_{x1} + n_{2}\phi_{x2} + n_{3}\phi_{x3} - n_{1}\phi_{x4} - n_{2}\phi_{x5} - n_{3}\phi_{x6} \rangle\rangle & \\  \nonumber
                                    ={}& \langle\langle  v_{n_{1}}^{2} v_{n_{2}}^{2} v_{n_{3}}^{2} \cos(n_{1}\psi_{n_{1}} + n_{2}\psi_{n_{2}} + n_{3}\psi_{n_{3}} -  n_{1}\psi_{n_{1}} -  n_{2}\psi_{n_{2}} -  n_{3}\psi_{n_{3}} )\rangle\rangle, & \\  \nonumber
                       ={}& \langle\langle  v^{2}_{n_{1}}  v^{2}_{n_{2}} v^{3}_{n_{3}} \rangle\rangle. &
\end{flalign}
}

\subsubsection{The k-particle asymmetric correlations:}

The three-, four- and five-particle asymmetric correlations can be given using the multi-particle correlation methods:
{
\begin{flalign}\label{eq:2-2}
    ASC(n_1, n_2, n_3)_{2-Sub} ={}& \langle\langle  \cos(n_{1}\phi^{a}_{x1} + n_{2}\phi^{a}_{x2} + n_{3}\phi^{b}_{x3} \rangle\rangle, & \\  \nonumber
                       ={}& \langle\langle  v_{n_{1}} v_{n_{2}} v_{n_{3}} \cos(n_{1}\psi_{n_{1}} + n_{2}\psi_{n_{2}} + n_{3}\psi_{n_{3}} )\rangle\rangle,  &
\end{flalign}
}
{
\begin{flalign}\label{eq:2-4}
   ASC(n_1, n_2, n_3, n_4)_{2-Sub} ={}& \langle\langle  \cos(n_{1}\phi^{a}_{x1} + n_{2}\phi^{a}_{x2} + n_{3}\phi^{b}_{x3} + n_{4}\phi^{b}_{x4} \rangle\rangle, & \\  \nonumber
                       ={}& \langle\langle  v_{n_{1}} v_{n_{2}} v_{n_{3}} v_{n_{4}} \cos(n_{1}\psi_{n_{1}} + n_{2}\psi_{n_{2}} + n_{3}\psi_{n_{3}} +  n_{4}\psi_{n_{4}} )\rangle\rangle,  &
\end{flalign}
}
{
\begin{flalign}\label{eq:2-5}
   ASC(n_1, n_2, n_3, n_4, n_5)_{Trad} ={}& \langle\langle  \cos(n_{1}\phi_{x1} + n_{2}\phi_{x2} + n_{3}\phi_{x3} + n_{4}\phi_{x4} + n_{5}\phi_{x5}\rangle\rangle, & \\  \nonumber
                       ={}& \langle\langle  v_{n_{1}} v_{n_{2}} v_{n_{3}} v_{n_{4}} v_{n_{5}} \cos(n_{1}\psi_{n_{1}} + n_{2}\psi_{n_{2}} + n_{3}\psi_{n_{3}} +  n_{4}\psi_{n_{4}} +  n_{5}\psi_{n_{5}})\rangle\rangle.  &
\end{flalign}
}

\subsubsection{The k-particle normalized symmetric correlations:}
The multi-particle correlations Eqs.~\ref{eq:2-1} and~\ref{eq:2-3} can be used to give the same and mix order flow harmonics NSC as~\cite{ALICE:2016kpq, ATLAS:2014ndd}:
{\color{black}{
\begin{flalign}\label{eq:3-4}
\gamma_{n_1,n_1,-n_1,-n_1} ={}& 2 - \dfrac{SC(n_1,n_1,-n_1,-n_1)_{2-Sub}}{SC(n_1,-n_1)_{2-Sub} SC(n_1,-n_1)_{2-Sub}}, &
\end{flalign}
}
}
{
\begin{flalign}\label{eq:3-4-1}
\beta_{n_1,n_2,-n_1,-n_2} ={}& \dfrac{SC(n_1,n_2,-n_1,-n_2)_{2-Sub}}{SC(n_1,-n_1)_{2-Sub} SC(n_2,-n_2)_{2-Sub}} -1. &
\end{flalign}
}
 {\color{black}Note that Eq.~\ref{eq:3-4} can also be written as, $ -C_{n}\lbrace 4\rbrace/C_{n}\lbrace 2\rbrace$ where $C_{n}\lbrace 2\rbrace$ and $C_{n}\lbrace 4\rbrace$ are the second- and fourth-order cumulant of the $v_{n}$ distribution~\cite{Giacalone:2017uqx,Bilandzic:2010jr,Bilandzic:2013kga,Giacalone:2016eyu,Jia:2012ve}.}

{\color{black}
The ratio $\gamma_{n,n,-n,-n}$ is a metric for the n$^{th}$ harmonic flow fluctuations.
In the absence of event-by-event fluctuations, $\gamma_{n,n,-n,-n}=1$ for the Bessel-Gaussian distribution of the $v_{2}$, while $0 < \gamma_{n,n,-n,-n} < 1$ quantifies the strength of the flow fluctuations~\cite{Voloshin:2007pc}. 
For $n=3$, $\gamma_{n,n,-n,-n}=0$ the $v_3$ has a Gaussian distribution, while $\gamma_{n,n,-n,-n} \neq 0$ quantifies the deviation from Gaussianity.
%
%
%
}
The ratio $\beta_{n,m,-n,-m}$ estimates the mixed harmonics flow correlations, (i) $\beta_{n,m,-n,-m} = 0.0$ represents the absence of the flow correlations, (ii) $\beta_{n,m,-n,-m} > 0.0$ denotes the mode-coupling strength between the $n$ and $m$ flow harmonics, and (iii) $\beta_{n,m,-n,-m} < 0.0$ gives the strength of anti-correlation between the $n$ and $m$ harmonics.


\subsubsection{The k-particle normalized asymmetric correlations:}
The multi-particle correlations Eqs.~\ref{eq:2-1}--\ref{eq:2-6} can define the Normalized Asymmetric Correlations, which gives the flow angle ($\psi_{n}$) correlations as~\cite{ATLAS:2014ndd, ALICE:2017fcd, STAR:2022vkx}:
{
\begin{flalign}\label{eq:3-1}
\rho_{n_1,n_2,n_3} ={}& \dfrac{ASC(n_1,n_2,n_3)_{2-Sub}}{\sqrt{|SC(n_1,n_2,-n_1,-n_2)_{2-Sub} SC(n_3, -n_3)_{2-Sub}|}}, \\  \nonumber
                   \sim{}& \langle \cos(n_1 \psi_{n_1} + n_2 \psi_{n_1} + n_3 \psi_{n_3}) \rangle. &
\end{flalign}
}
{
\begin{flalign}\label{eq:3-2}
\rho_{n_1,n_2,n_3,n_4} ={}& \dfrac{ASC(n_1,n_2,n_3,n_4)_{2-Sub}}{\sqrt{|SC(n_1,n_2,n_3,-n_1,-n_2,-n_3)_{Trad} SC(n_4, -n_4)_{2-Sub}|}}, & \\  \nonumber
                         \sim{}& \langle \cos(n_1 \psi_{n_1} + n_2 \psi_{n_1} + n_3 \psi_{n_3} + n_4 \psi_{n_4}) \rangle. &
\end{flalign}
}
{
\begin{flalign}\label{eq:3-3}
\rho_{n_1,n_2,n_3,n_4,n_5} ={}& \dfrac{ASC(n_1,n_2,n_3,n_4,n_5)_{Trad}}{\sqrt{|SC(n_1,n_2,n_3,-n_1,-n_2,-n_3)_{Trad} SC(n_4,n_5,-n_4,-n_5)_{2-Sub}|}}, &\\  \nonumber
                         \sim{}& \langle \cos(n_1 \psi_{n_1} + n_2 \psi_{n_1} + n_3 \psi_{n_3} + n_4 \psi_{n_4} + n_5 \psi_{n_5}) \rangle. &
\end{flalign}
}
%
The definitions given in Eqs.~\ref{eq:3-1}, \ref{eq:3-2}, and~\ref{eq:3-3} are in line with the definition used by the ALICE and STAR experiments~\cite{ALICE:2017fcd, STAR:2022vkx}.

\end{widetext}

\section{Results and discussion}\label{sec:3}
This section will discuss the prediction of the k-particle symmetric (asymmetric) correlations and the flow harmonics magnitude and angular correlations in the isobaric collisions for different initial state configurations given in Tab~\ref{tab:1}. Here, it's important to note that the non-flow effects on the presented results have been investigated in our previous study Ref~\cite{Magdy:2022ize}.

\subsection{Symmetric Correlations}
\begin{figure}[!h] 
\includegraphics[width=1.0  \linewidth, angle=-0,keepaspectratio=true,clip=true]{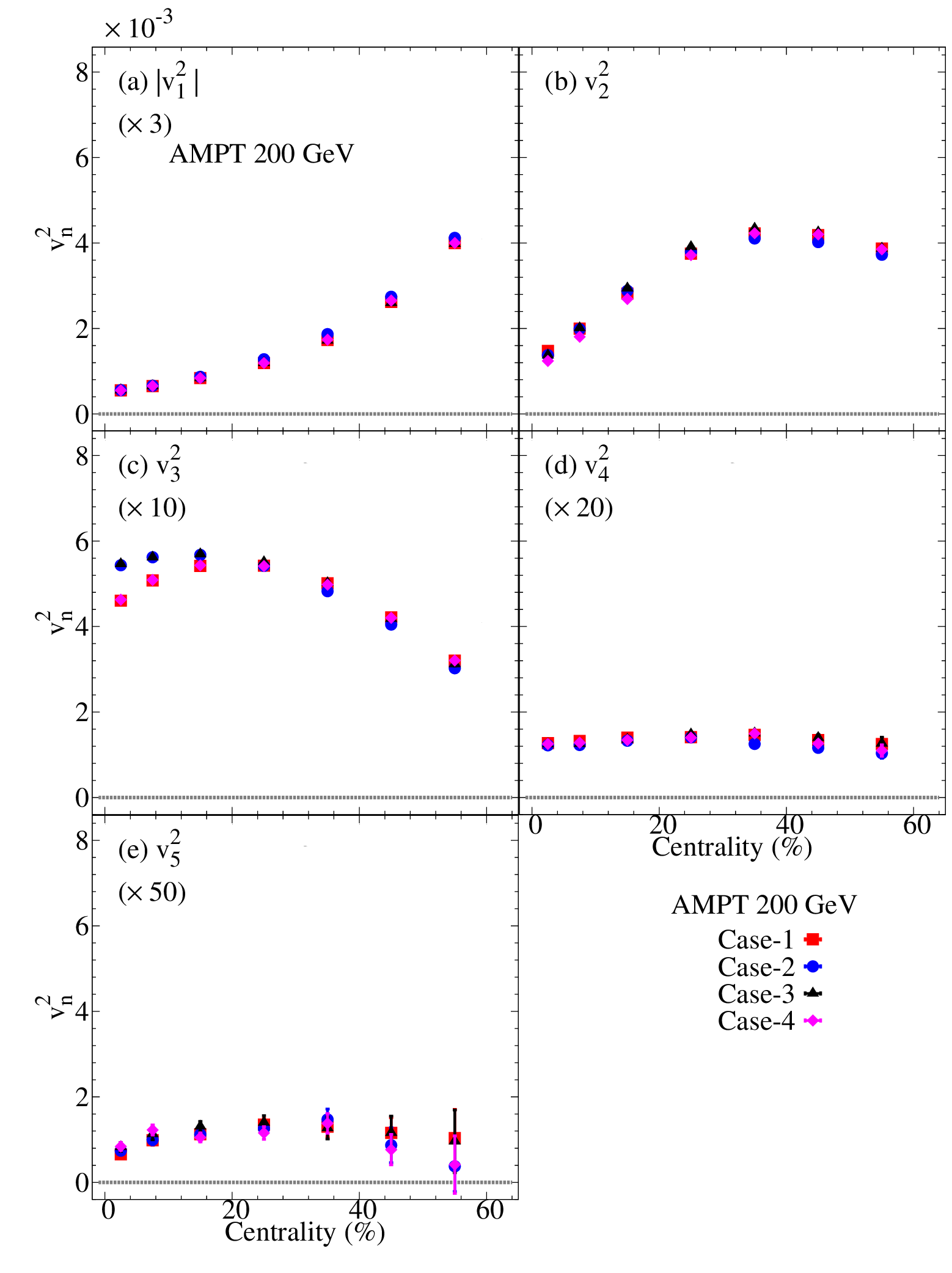}
\vskip -0.4cm
\caption{
Comparisons of the centrality dependence of the two-particle flow harmonics $v^{2}_{n} = SC$($n$,$-n$) using the two-subevents method for \RuRu and \ZrZr at $\sqrt{\textit{s}_{NN}}~=$ 200~GeV from the AMPT model with various parameters given in Tab.~\ref{tab:1}.
}\label{fig:c1}
\vskip -0.3cm
\end{figure}

\begin{figure}[!h] 
\includegraphics[width=1.0 \linewidth, angle=-0,keepaspectratio=true,clip=true]{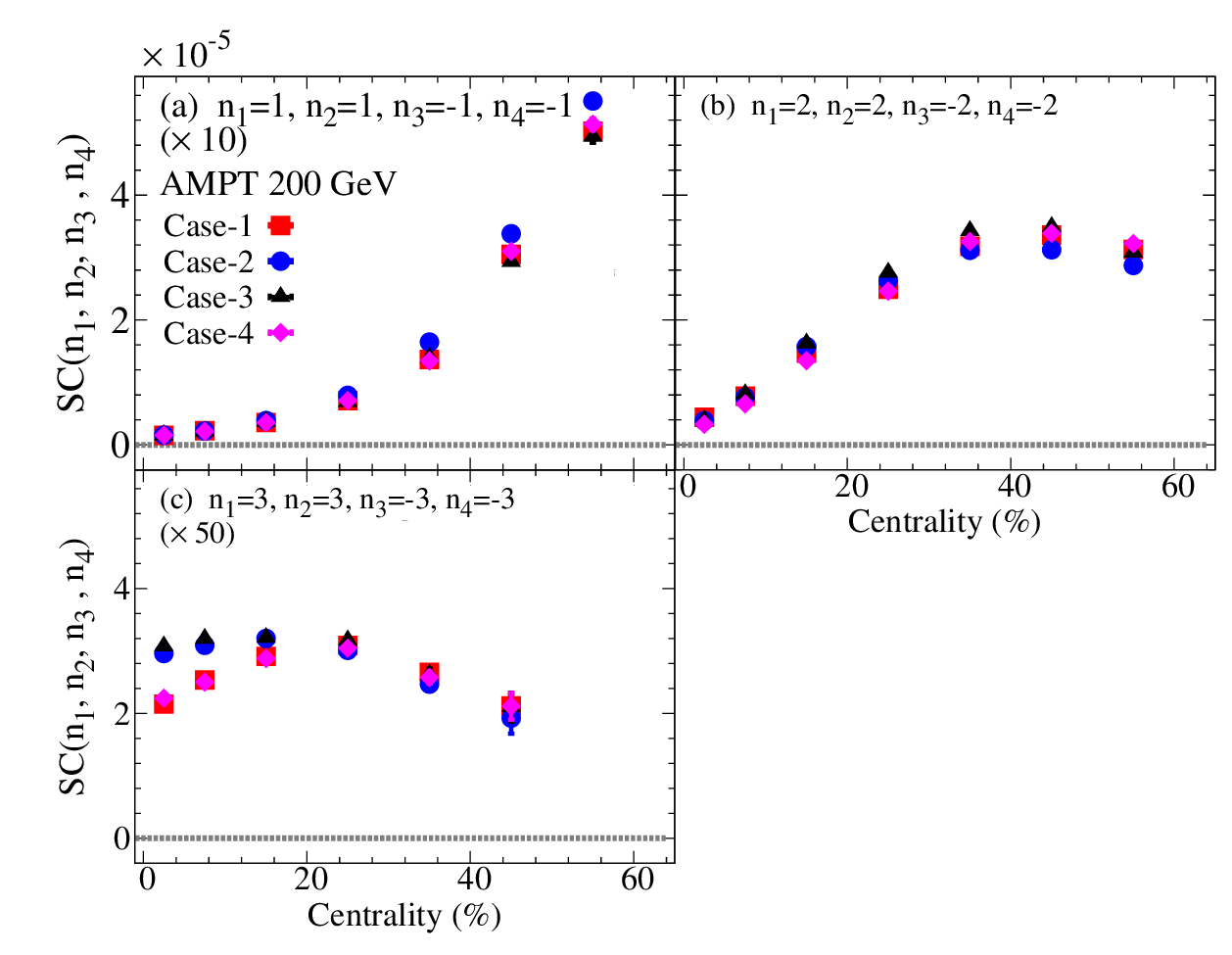}
\vskip -0.4cm
\caption{
Same as in Fig.~\ref{fig:c1} but for the four-particle $SC$($1$,$1$,$-1$,$-1$) panel (a),  $SC$($2$,$2$,$-2$,$-2$) panel (b) and  $SC$($3$,$3$,$-3$,$-3$) panel (c) using the two-subevents method.
}\label{fig:c2}
\vskip -0.3cm
\end{figure}

\begin{figure}[!h] 
\includegraphics[width=1.0 \linewidth, angle=-0,keepaspectratio=true,clip=true]{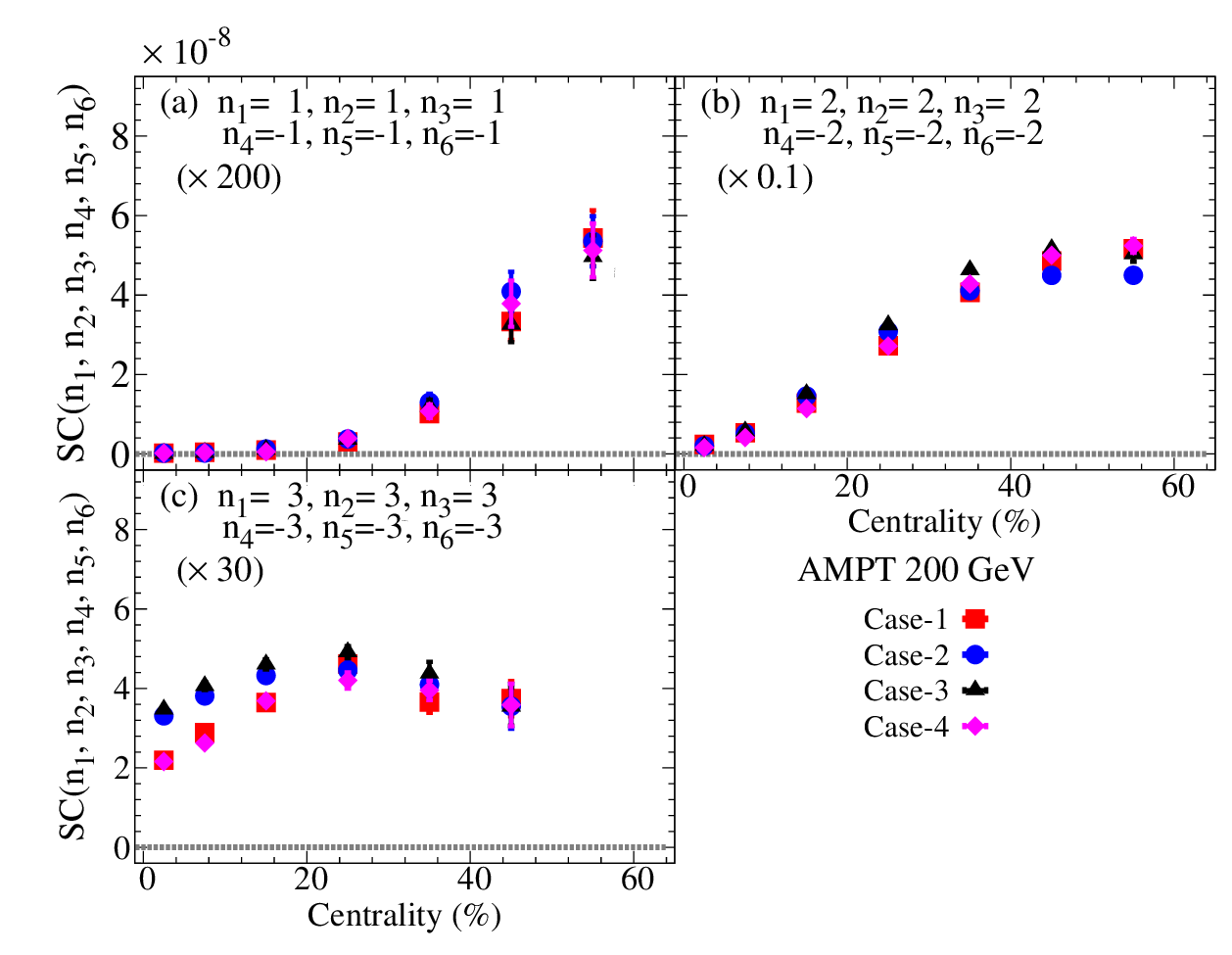}
\vskip -0.4cm
\caption{
Same as in Fig.~\ref{fig:c1} but for the six-particle $SC$($1$,$1$,$1$,$-1$,$-1$,$-2$) panel (a),  $SC$($2$,$2$,$2$,$-2$,$-2$,$-2$) panel (b) and  $SC$($3$,$3$,$3$,$-3$,$-3$,$-3$) panel (c), using the one-subevent method.
}\label{fig:c3}
\vskip -0.3cm
\end{figure}

Fugues~\ref{fig:c1},~\ref{fig:c2}, and~\ref{fig:c3} show the centrality and system size dependencies of the two-, four-, and six-particle symmetric correlations. The comparisons of the same flow harmonic two-, four-, and six-particle correlations are expected to reflect the initial state density fluctuations and the final state hydrodynamic evolution fluctuations~\cite{STAR:2022gki}.
The same harmonic correlations (n$>$1) from the AMPT model at the same $\sigma_{pp}$ present characteristic patterns from central to peripheral collisions (i.e., increasing from central to mid-central collisions, then decreasing as the collisions become more peripheral). Such characteristic patterns display the interplay between initial-state eccentricities and final-state viscous attenuation~\cite{Lacey:2013is}. Note that the n$=$1 correlations will be impacted by the global momentum conservation (GMC) effect~\cite{STAR:2018gji, STAR:2019zaf}, which have been recently studied in the isobaric collisions~\cite{Magdy:2023fsp}.

\begin{figure}[!h] 
\includegraphics[width=1.0 \linewidth, angle=-0,keepaspectratio=true,clip=true]{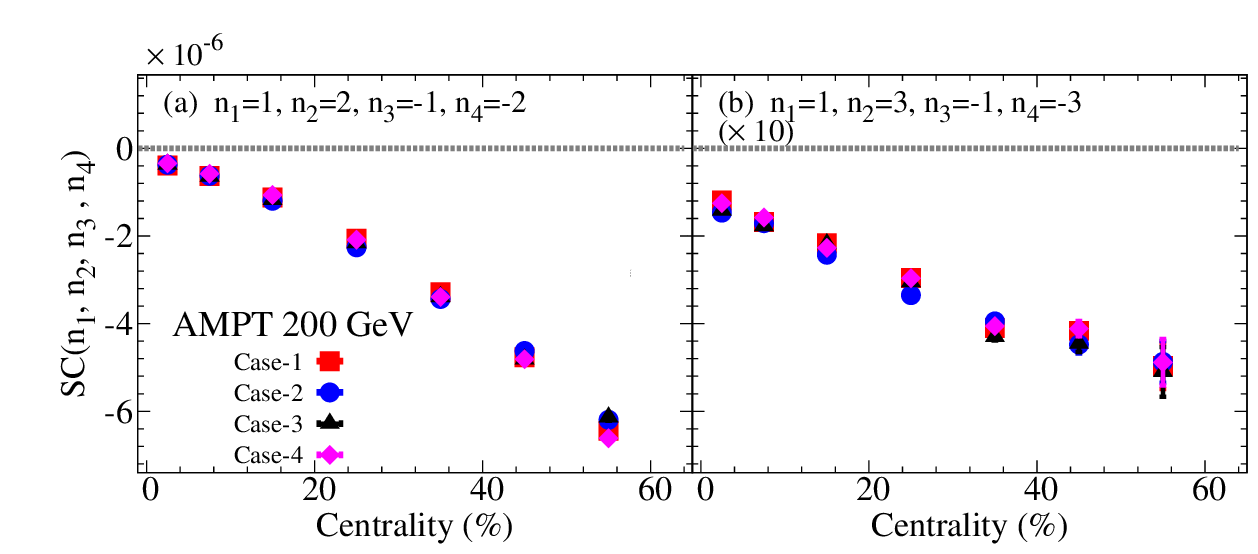}
\vskip -0.4cm
\caption{
Same as in Fig.~\ref{fig:c1} but for the four-particle $SC$($1$,$2$,$-1$,$-2$) panel (a) and $SC$($1$,$3$,$-1$,$-3$) panel (b), using the two-subevent method.
}\label{fig:c4}
\vskip -0.3cm
\end{figure}
\begin{figure}[!h] 
\includegraphics[width=1.0 \linewidth, angle=-0,keepaspectratio=true,clip=true]{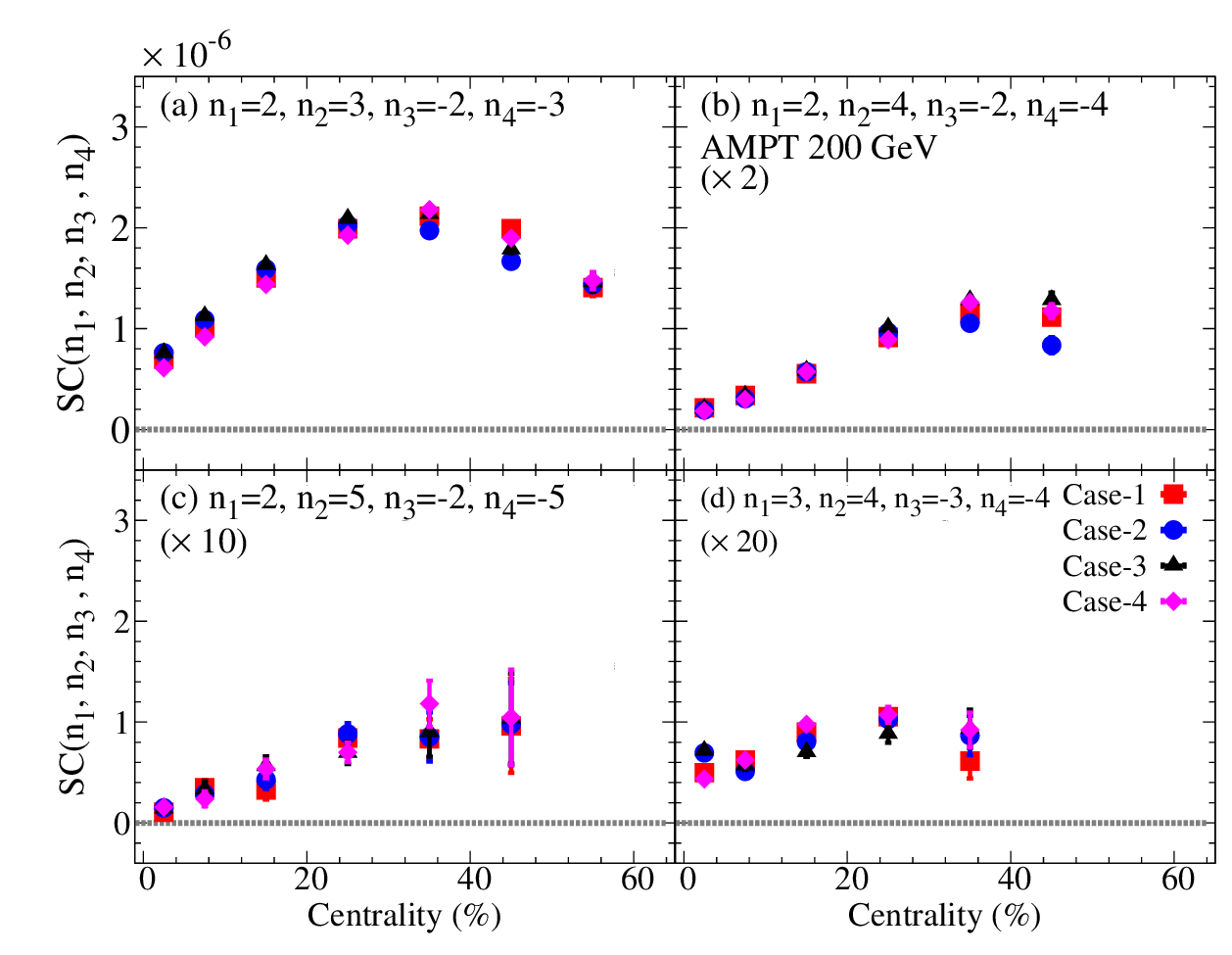}
\vskip -0.4cm
\caption{
Same as in Fig.~\ref{fig:c1} but for the four-particle $SC$($2$,$3$,$-2$,$-3$) panel (a), $SC$($2$,$4$,$-2$,$-4$) panel (b), $SC$($2$,$4$,$-2$,$-4$) panel (c), and $SC$($3$,$4$,$-3$,$-4$) panel (d), using the two-subevent method.
}\label{fig:c5}
\vskip -0.3cm
\end{figure}
\begin{figure}[!h] 
\includegraphics[width=1.0 \linewidth, angle=-0,keepaspectratio=true,clip=true]{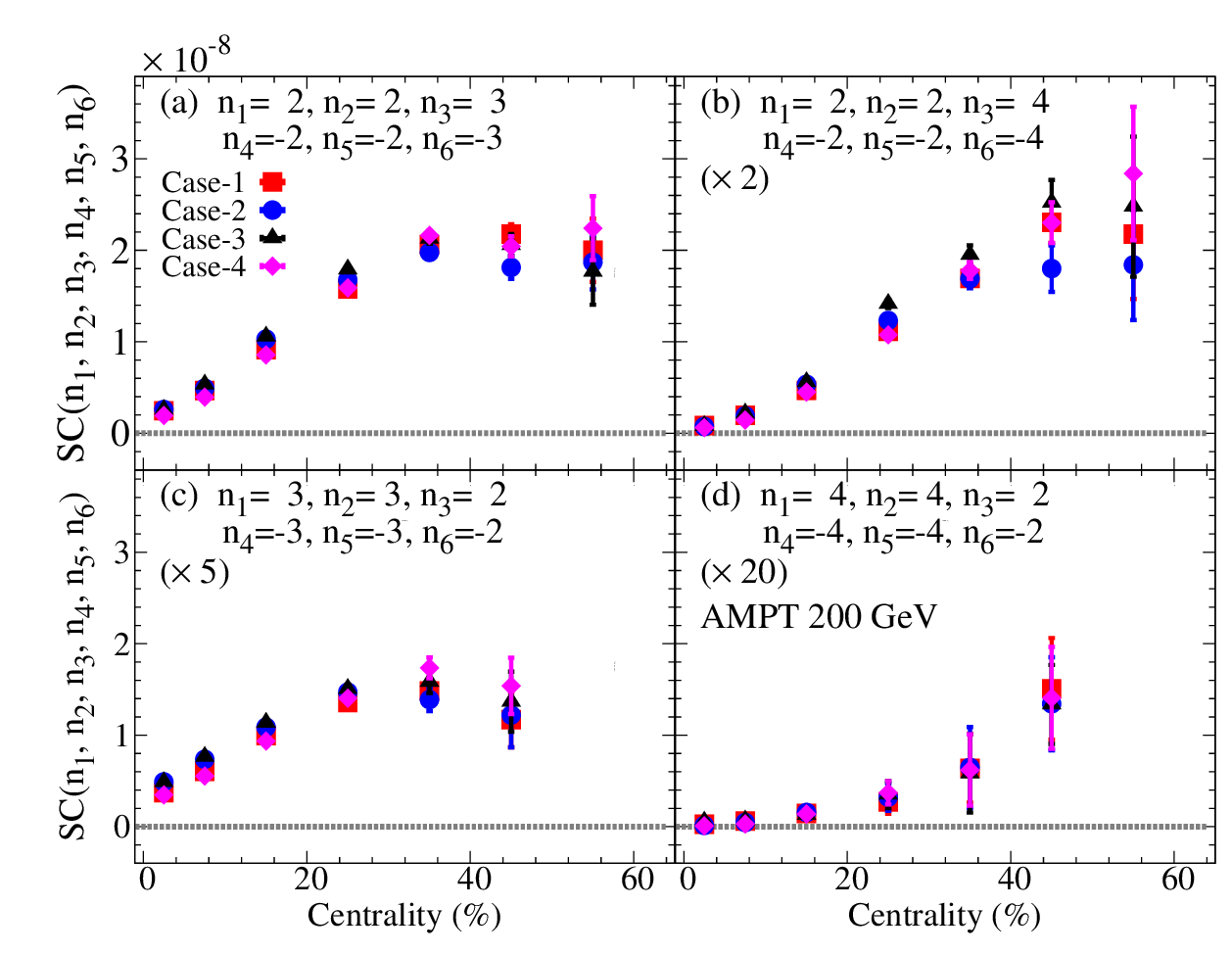}
\vskip -0.4cm
\caption{
Same as in Fig.~\ref{fig:c1} but for the six-particle symmetric correlations $SC$($2$,$2$,$3$,$-2$,$-2$,$-3$) panel (a),  $SC$($2$,$2$,$4$,$-2$,$-2$,$-4$) panel (b),  $SC$($3$,$3$,$2$,$-3$,$-3$,$-2$) panel (c), and $SC$($4$,$4$,$2$,$-4$,$-4$,$-2$) panel (d),  using the one-subevent method.
}\label{fig:c6}
\vskip -0.3cm
\end{figure}
\begin{figure}[!h] 
\includegraphics[width=1.0 \linewidth, angle=-0,keepaspectratio=true,clip=true]{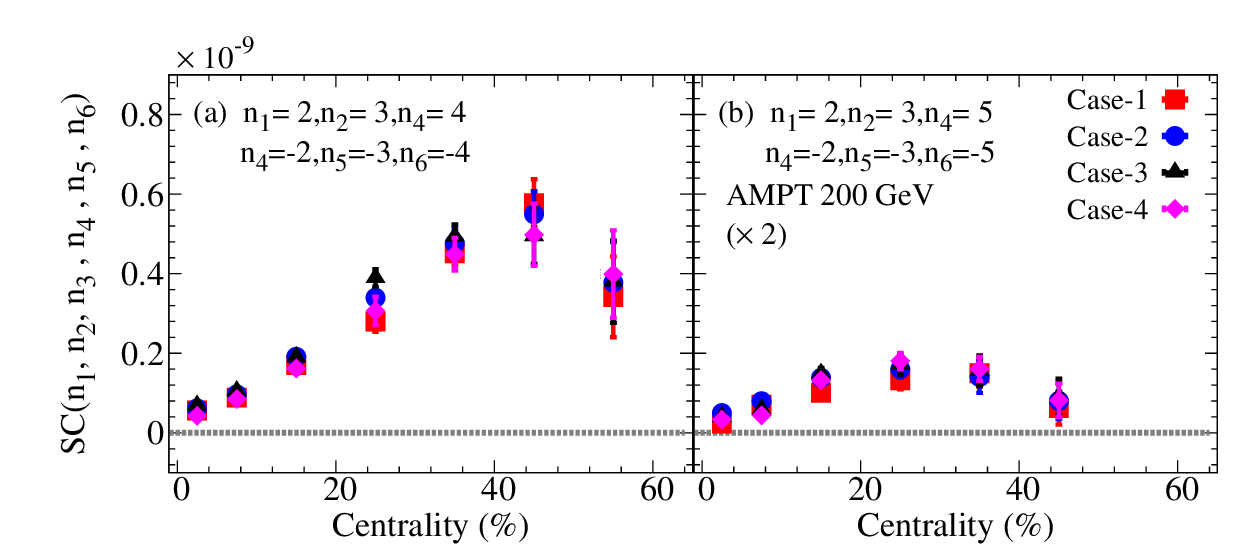}
\vskip -0.4cm
\caption{
Same as in Fig.~\ref{fig:c1} but for the six-particle symmetric correlations $SC$($2$,$3$,$4$,$-2$,$-3$,$-4$) panel (a) and  $SC$($2$,$3$,$5$,$-2$,$-3$,$-5$) panel (b), using the one-subevent method.
}\label{fig:c7}
\vskip -0.3cm
\end{figure}

The four- and six-particle mixed harmonics correlations as a function of centrality for \RuRu and \ZrZr at $\sqrt{\textit{s}_{NN}}~=$ 200~GeV from the AMPT model with various parameters given in Tab.~\ref{tab:1} are presented in Figs.~\ref{fig:c4},~\ref{fig:c5},~\ref{fig:c6}, and~\ref{fig:c7}. These mixed harmonics flow correlations are expected to give the flow harmonics magnitude correlations induced by initial and final state effects, which will be provided employing the ratios defined in Eq.~\ref{eq:3-4-1}. The correlations presented in  Figs.~\ref{fig:c4},~\ref{fig:c5},~\ref{fig:c6}, and~\ref{fig:c7} are sensitive to the interplay between final and initial state effects~\cite{Magdy:2022ize}. The mixed harmonics correlations with $n=1$, Fig.~\ref{fig:c4}, are expected to be impacted by the GMC effect.
\subsection{Asymmetric correlations}
Using the symmetric and asymmetric correlations, we can get the flow angular correlations induced by the initial state and given by Eqs.~\ref{eq:3-1}--\ref{eq:3-3}. Therefore, it's constructive first to discuss the centrality and system size dependence of asymmetric correlations. 
\begin{figure}[!h] 
\includegraphics[width=1.0 \linewidth, angle=-0,keepaspectratio=true,clip=true]{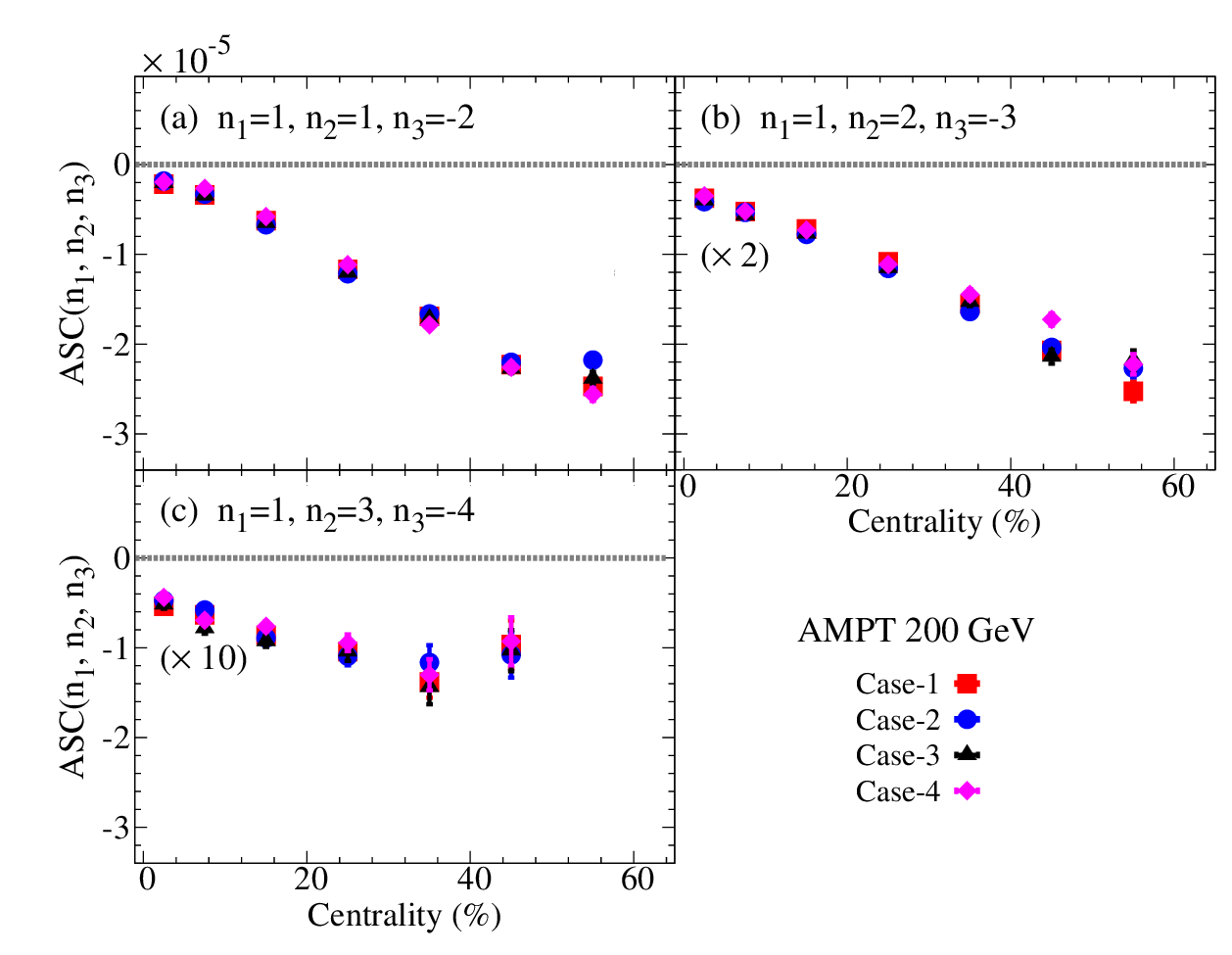}
\vskip -0.4cm
\caption{
Comparisons of the centrality dependence of the three-particle asymmetric correlations $ASC$($1$,$1$,$-2$) panel (a),  $ASC$($1$,$2$,$-3$) panel (b) and  $ASC$($1$,$3$,$-4$) panel (c), using the two-subevents method for \RuRu and \ZrZr at $\sqrt{\textit{s}_{NN}}~=$ 200~GeV from the AMPT model with various parameters given in Tab.~\ref{tab:1}.
}\label{fig:c8}
\vskip -0.3cm
\end{figure}
\begin{figure}[!h] 
\includegraphics[width=1.0 \linewidth, angle=-0,keepaspectratio=true,clip=true]{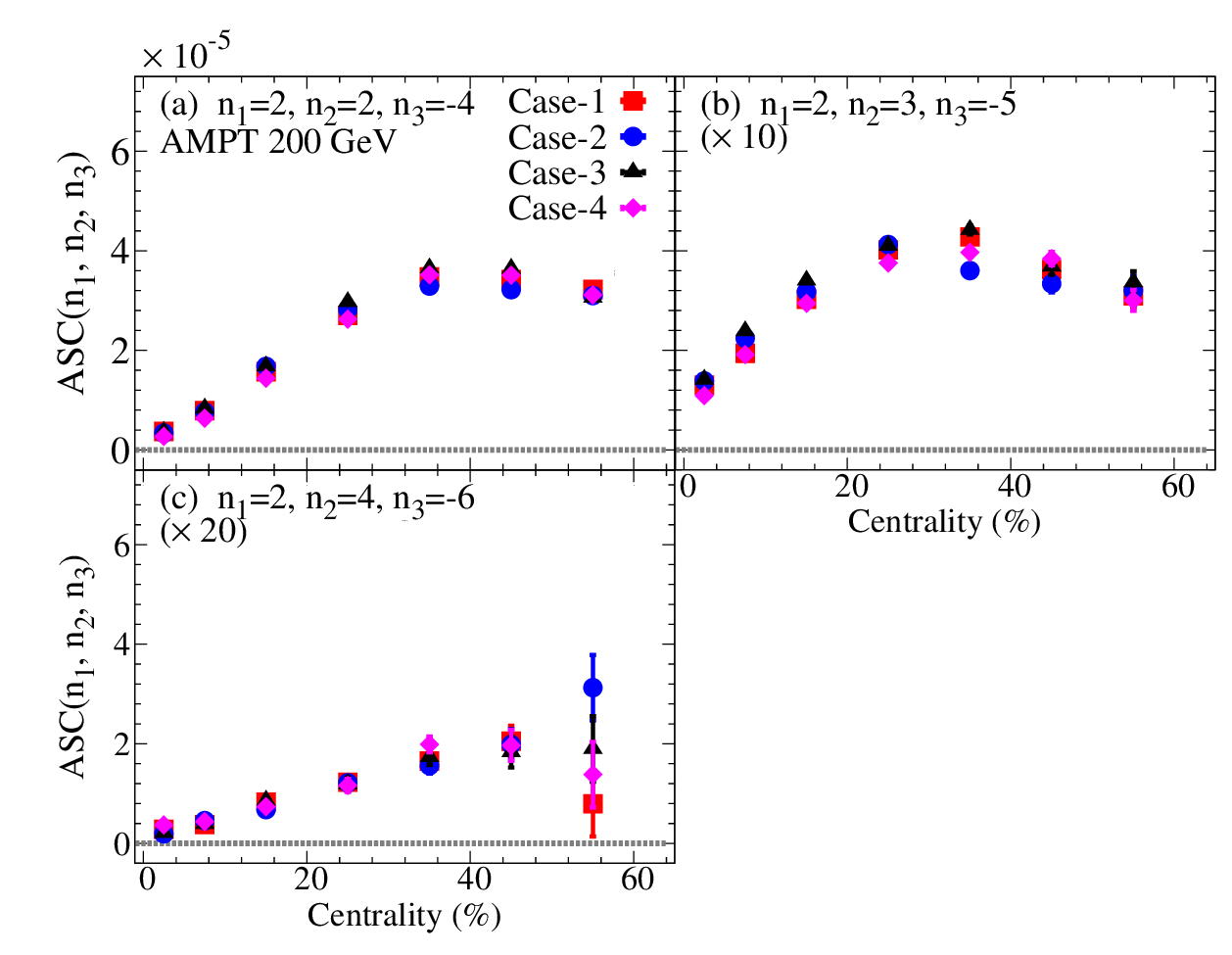}
\vskip -0.4cm
\caption{
Same as in Fig.~\ref{fig:c8} but for the three-particle asymmetric correlations $ASC$($2$,$2$,$-4$) panel (a),  $ASC$($2$,$3$,$-5$) panel (b) and  $ASC$($2$,$4$,$-5$) panel (c), using two-subevents method.
}\label{fig:c9}
\vskip -0.3cm
\end{figure}
\begin{figure}[!h] 
\includegraphics[width=1.0 \linewidth, angle=-0,keepaspectratio=true,clip=true]{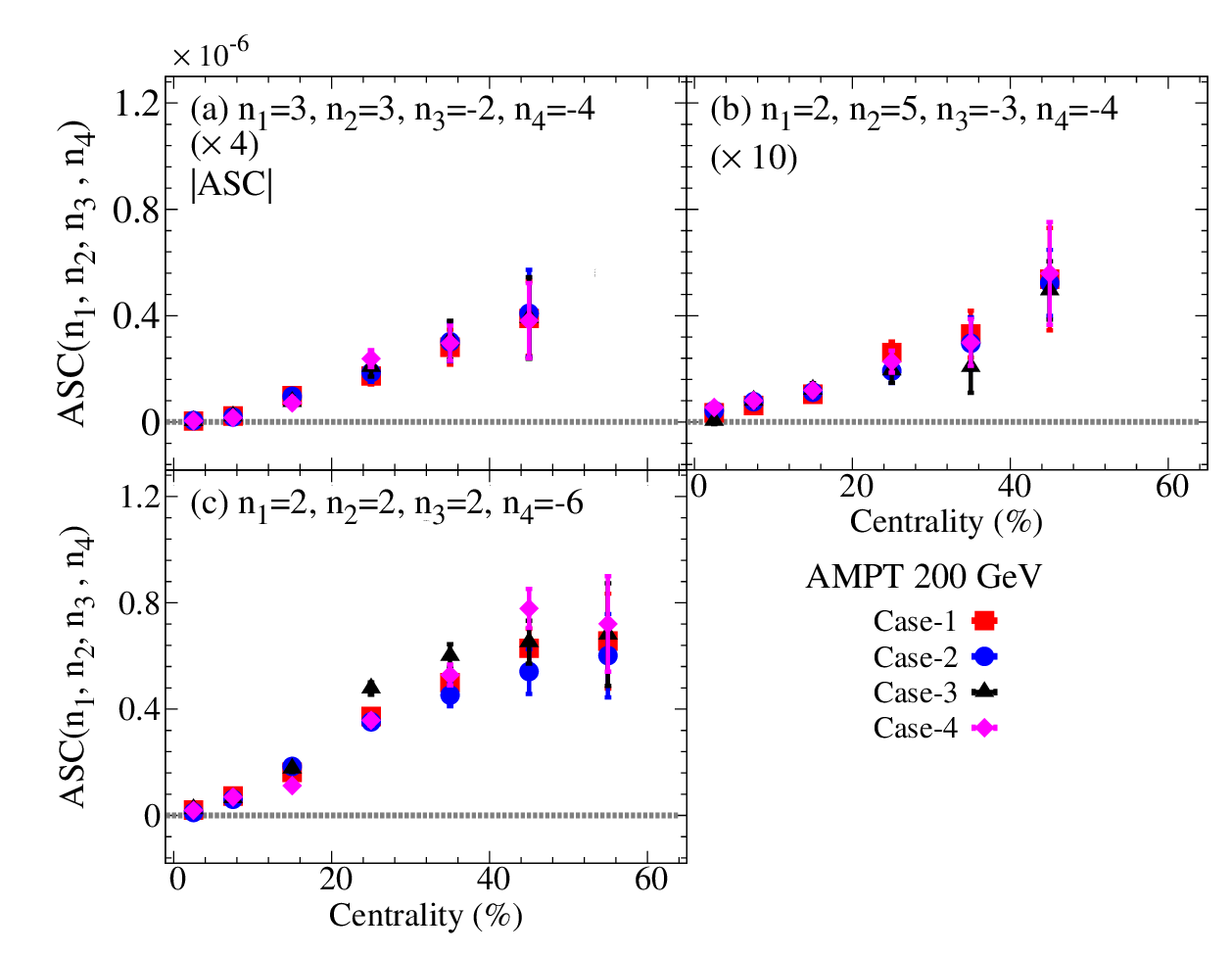}
\vskip -0.4cm
\caption{
Same as in Fig.~\ref{fig:c8} but for the four-particle asymmetric correlations $ASC$($3$,$3$,$-2$,$-4$) panel (a),  $ASC$($2$,$5$,$-3$,$-4$) panel (b) and  $ASC$($2$,$2$,$2$,$-6$) panel (c), using two-subevents method.
}\label{fig:c10}
\vskip -0.3cm
\end{figure}
\begin{figure}[!h] 
\includegraphics[width=1.0 \linewidth, angle=-0,keepaspectratio=true,clip=true]{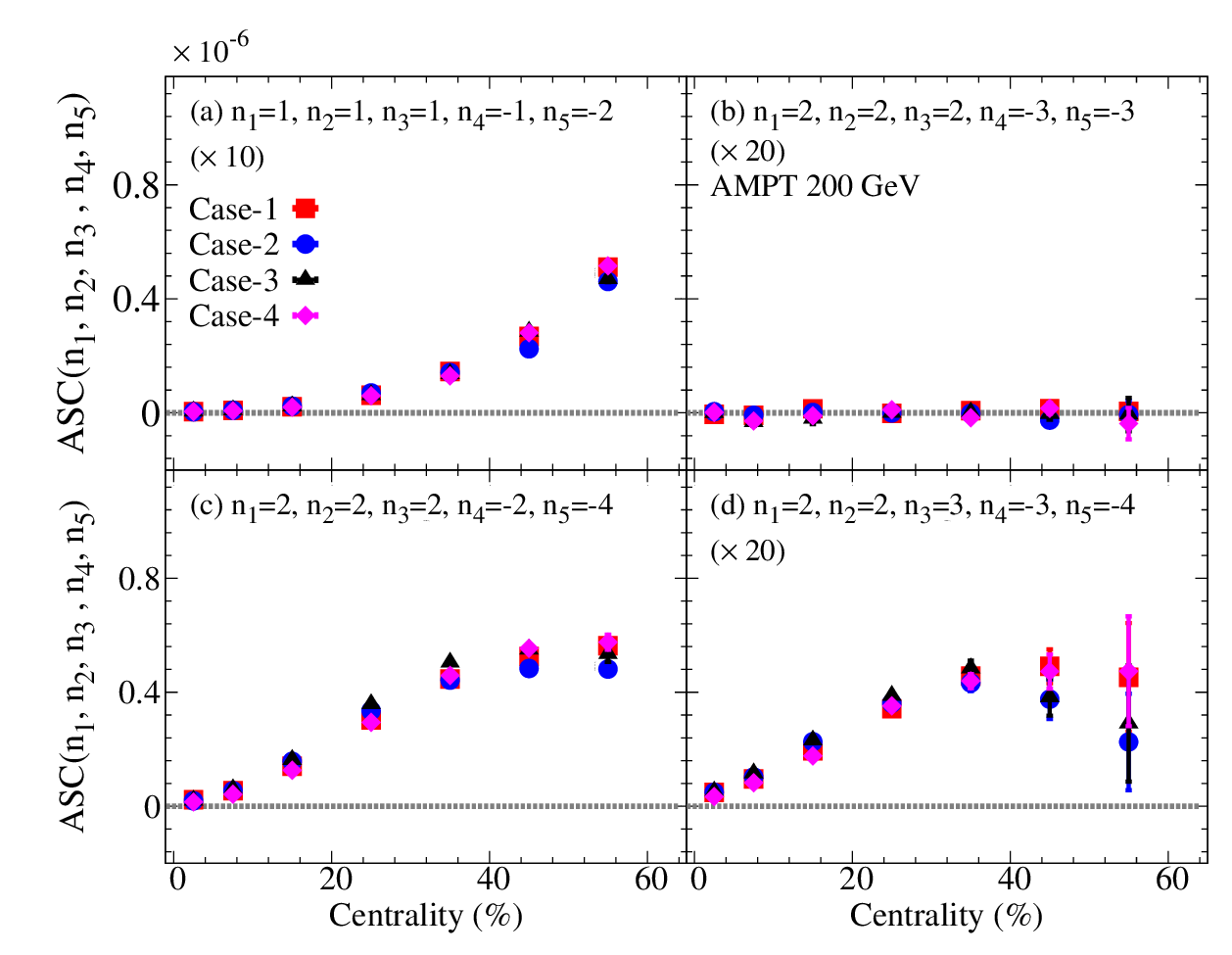}
\vskip -0.4cm
\caption{
Same as in Fig.~\ref{fig:c8} but for the five-particle asymmetric correlations $ASC$($1$,$1$,$1$,$-1$,$-2$) panel (a),  $ASC$($2$,$2$,$2$,$-3$,$-3$) panel (b), $ASC$($2$,$2$,$2$,$-2$,$-4$) panel (c), and $ASC$($2$,$2$,$3$,$-3$,$-4$) panel (d), using one-subevent method.
}\label{fig:c11}
\vskip -0.3cm
\end{figure}

The three-, four-, and five-particle asymmetric correlations for \RuRu and \ZrZr at $\sqrt{\textit{s}_{NN}}~=$ 200~GeV from the AMPT model are shown in Figs.~\ref{fig:c8}--~\ref{fig:c11}. 
The presented AMPT calculations of the ASC Figs.~\ref{fig:c8}--~\ref{fig:c11} show that the strength of the correlation gets stronger as the collisions become more peripheral. In contrast, the $ASC$($2$,$2$,$2$,$-3$,$-3$) values in Fig.~\ref{fig:c11} panel (b) are consistent with zero, reflecting the weak correlation nature between $\psi_{2}$ and $\psi_{3}$. 

\subsection{Normalized symmetric correlations}
In our previous studies~\cite{Magdy:2023owx, Magdy:2022ize}, we pointed out that the flow harmonics magnitude fluctuations and correlations have a weak sensitivity to the final state effects. Therefore, they can constrain the HIC initial conditions~\cite{STAR:2022gki, STAR:2022vkx}. The flow harmonics magnitude fluctuations and correlations can be given $\gamma_{n,n,-n,-n}$ Eq.~\ref{eq:3-4} and $\beta_{n,m,-n,-m}$ Eq.~\ref{eq:3-4-1} respectively. 
For NSC, we will present the AMPT model calculations for Case-1 and Case-2 parameters see Tab.~\ref{tab:1}.

\begin{figure}[!h] 
\includegraphics[width=1.0 \linewidth, angle=-0,keepaspectratio=true,clip=true]{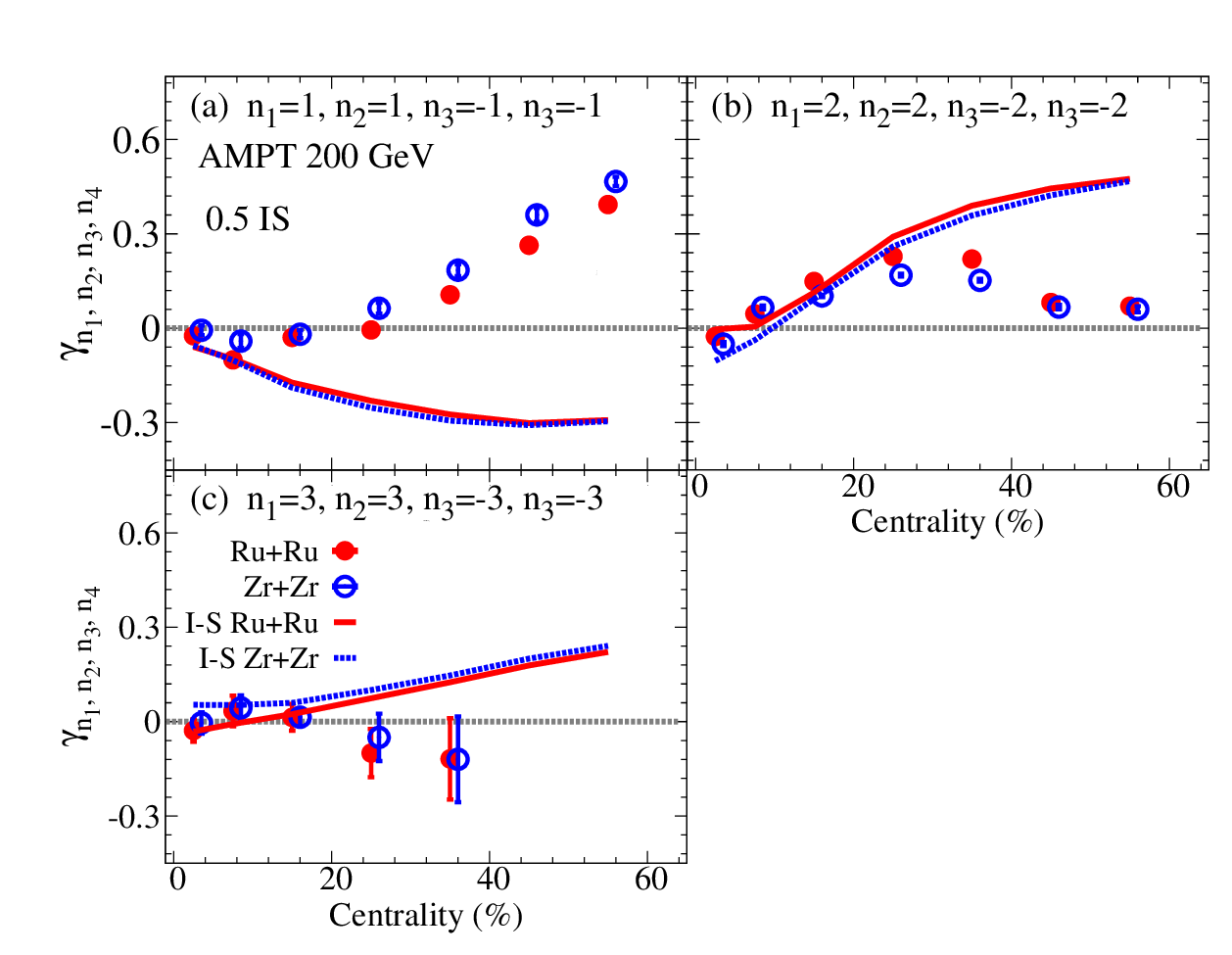}
\vskip -0.4cm
\caption{
Comparisons of the centrality dependence of the same harmonic normalized symmetric correlations $\gamma_{1,1,-1,-1}$ panel (a), $\gamma_{2,2,-2,-2}$ panel (b) and $\gamma_{3,3,-3,-3}$ panel (c), for \RuRu and \ZrZr at $\sqrt{\textit{s}_{NN}}~=$ 200~GeV from the AMPT model Case-1 and Case-2 parameters given in Tab.~\ref{tab:1}. The curves represent the initial state eccentricity fluctuations.
}\label{fig:r1}
\vskip -0.3cm
\end{figure}
\begin{figure}[!h] 
\includegraphics[width=1.0 \linewidth, angle=-0,keepaspectratio=true,clip=true]{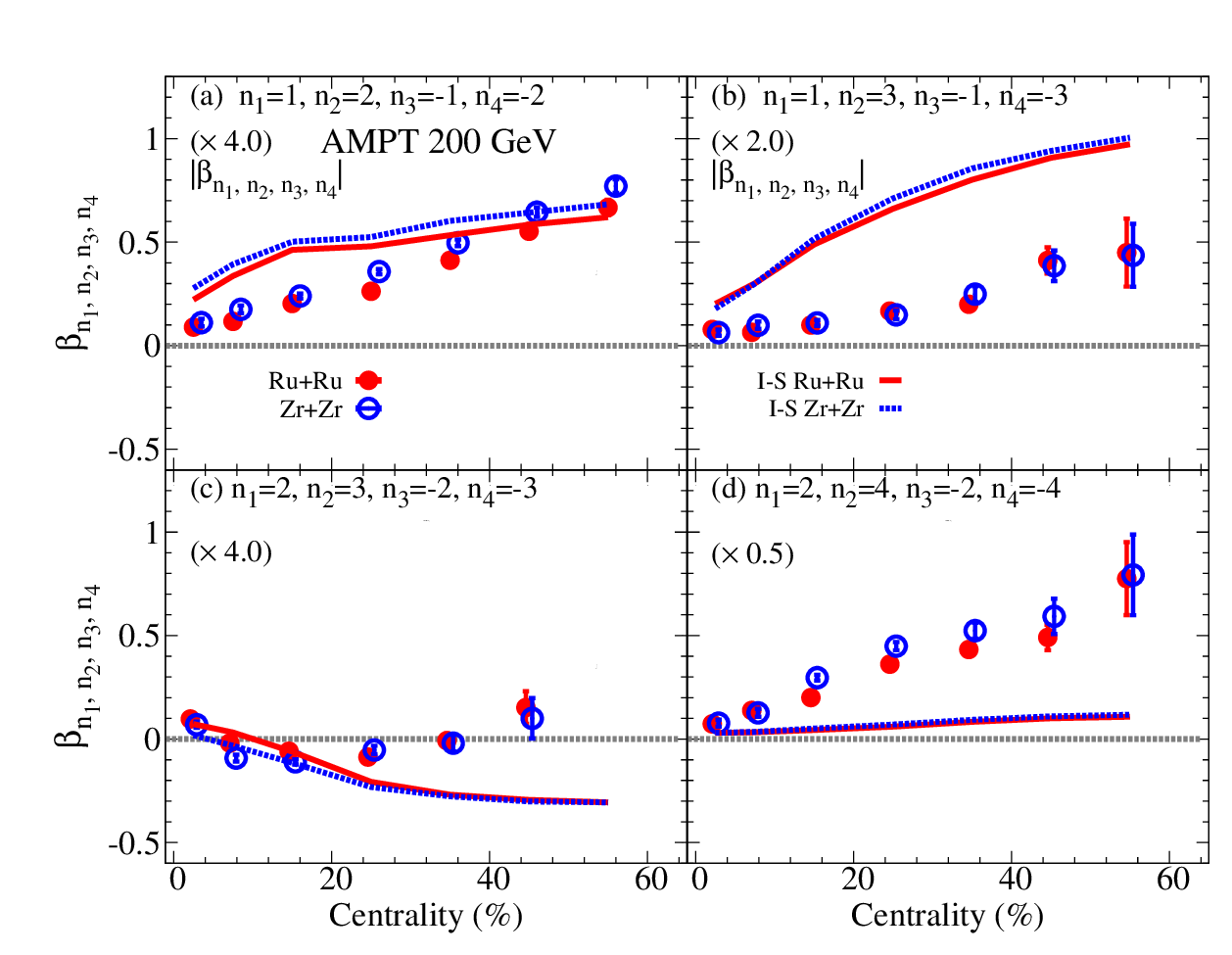}
\vskip -0.4cm
\caption{
Same as in Fig.~\ref{fig:r1} but for the mixed harmonic normalized symmetric correlations $\beta_{1,2,-1,-2}$ panel (a), $\beta_{1,3,-1,-3}$ panel (b), $\beta_{2,3,-2,-3}$ panel (c), $\beta_{2,4,-2,-4}$ panel (d), $\beta_{2,5,-2,-5}$ panel (e), and $\beta_{3,4,-3,-4}$ panel (f).
}\label{fig:r2}
\vskip -0.3cm
\end{figure}

The normalized symmetric correlations for the same flow harmonic, $\gamma_{1,1,-1,-1}$, $\gamma_{2,2,-2,-2}$ and $\gamma_{3,3,-3,-3}$ are presented in Fig.~\ref{fig:r1} for \RuRu and \ZrZr at $\sqrt{\textit{s}_{NN}}~=$ 200~GeV from the AMPT model Case-1 and Case-2. 
In panel (a), the $\gamma_{1,1,-1,-1}$ indicates the rapidity-even dipolar flow fluctuations in the AMPT model. The $\gamma_{1,1,-1,-1}$ calculations assume equivalent GMC effects on the $n=1$ two- and four-particle correlations. Our calculations indicated an apparent difference in the fluctuation nature between the $\gamma_{1,1,-1,-1}$ calculated for initial and final states in the AMPT model. 
The elliptic flow fluctuations are given in panel (b) by the $\gamma_{2,2,-2,-2}$; it indicates the anticipated reduction in the fluctuations magnitude from central to peripheral collisions. We observe an agreement between $\gamma_{2,2,-2,-2}$ calculated for initial and final states in the AMPT model in central collisions. In contrast, we observed less fluctuations in the initial state $\gamma_{2,2,-2,-2}$. The latter observation suggests that the AMPT final state effects add more fluctuations to the elliptic flow.
{\color{black}The ratio $\gamma_{3,3,-3,-3}$ in panel (c) presents the triangular flow fluctuations. The $\gamma_{3,3,-3,-3}$ from the AMPT model is consistent with zero with significant uncertainties for both systems, which is consistent with the STAR measurements~\cite{STAR:2013qio} and the prior AMPT calculations ~\cite{Magdy:2022ize} for Au+Au at 200 GeV.}

Figure~\ref{fig:r2} show the mixed flow harmonics NSC $\beta_{1,2,-1,-2}$ panel (a), $\beta_{1,3,-1,-3}$ panel (b), $\beta_{2,3,-2,-3}$ panel (c) and $\beta_{2,4,-2,-4}$ panel (d) for \RuRu and \ZrZr at $\sqrt{\textit{s}_{NN}}~=$ 200~GeV from the AMPT model Case-1 and Case-2. 
The correlations between the rapidity-even dipolar flow $v_1$ and the $v_2$ and $v_3$ magnitudes given in panels (a) and (b) suppose equal GMC effects on the $n=1$ two- and four-particle correlations. The $\beta_{1,2,-1,-2}$ and $\beta_{1,3,-1,-3}$ initial and final state calculations indicated anticorrelations between $v_1$ and $v_n$ ($n$=2,3).
On the other hand, the $v_2$--$v_n$ (n$=$3 and 4) correlations nature are given by $\beta_{2,3,-2,-3}$, $\beta_{2,4,-2,-4}$ in panels (c) and (d). The initial and final state calculations show anticorrelations between $v_2$ and $v_3$, with larger strength for the initial state calculations. In contrast, we observed positive correlations between $v_2$ and $v_4$ from initial and final state results, with much smaller strength for the initial state calculations.

\subsection{Normalized asymmetric correlations}
An additional understanding of the HIC initial conditions~\cite{STAR:2022vkx} can be gained by studying the flow angular correlations via the normalized asymmetric correlations Eqs.~\ref{eq:3-1}--\ref{eq:3-3}. The NASC $\rho_{X}$ is expected to operate as a metric for the strength of the correlations between flow symmetry planes (i.e., $\psi_{1}$--$\psi_{5}$) see Tab\ref{tab:2}.
\begin{table}[h!]
\begin{center}
\caption{The summary of the NASC that is presented in this work.\label{tab:2}}
 \begin{tabular}{|c|c|c|c|}
 \hline 
   Event-plane angular correlations   &  NASC   \\
  \hline
  $\langle \cos(2 \psi_{1} - 2 \psi_{2}) \rangle$                             & $\rho_{1,1,-2}$        \\
  $                                             $                             & $\rho_{1,1,1,-1,-2}$   \\
  \hline
  $\langle \cos(6 \psi_{2} - 6 \psi_{3}) \rangle$                             & $\rho_{2,2,2,-3,-3}$    \\
  \hline
  $                                             $                             & $\rho_{2,2,-4}$         \\
  $\langle \cos(4 \psi_{2} - 4 \psi_{4}) \rangle$                             & $\rho_{2,2,2,-2,-4}$    \\
  $                                             $                             & $\rho_{2,2,3,-3,-4}$    \\
  \hline
  $\langle \cos(6 \psi_{2} - 6 \psi_{6}) \rangle$                             & $\rho_{2,2,2,-6}$      \\
  \hline
  $\langle \cos(1 \psi_{1} + 2 \psi_{2} - 3 \psi_{3}) \rangle$                & $\rho_{1,2,-3}$        \\
  \hline
  $\langle \cos(1 \psi_{1} + 3 \psi_{3} - 4 \psi_{4} \rangle$                 & $\rho_{1,3,-4}$        \\
  \hline
  $\langle \cos(2 \psi_{2} + 3 \psi_{3} - 5 \psi_{5}) \rangle$                & $\rho_{2,3,-5}$        \\
  \hline
  $\langle \cos(6 \psi_{3} - 2 \psi_{2} - 4 \psi_{4}) \rangle$                & $\rho_{3,3,-2,-4}$     \\
  \hline
\end{tabular}
\end{center}
\end{table}

\begin{figure}[!h] 
\includegraphics[width=1.0 \linewidth, angle=-0,keepaspectratio=true,clip=true]{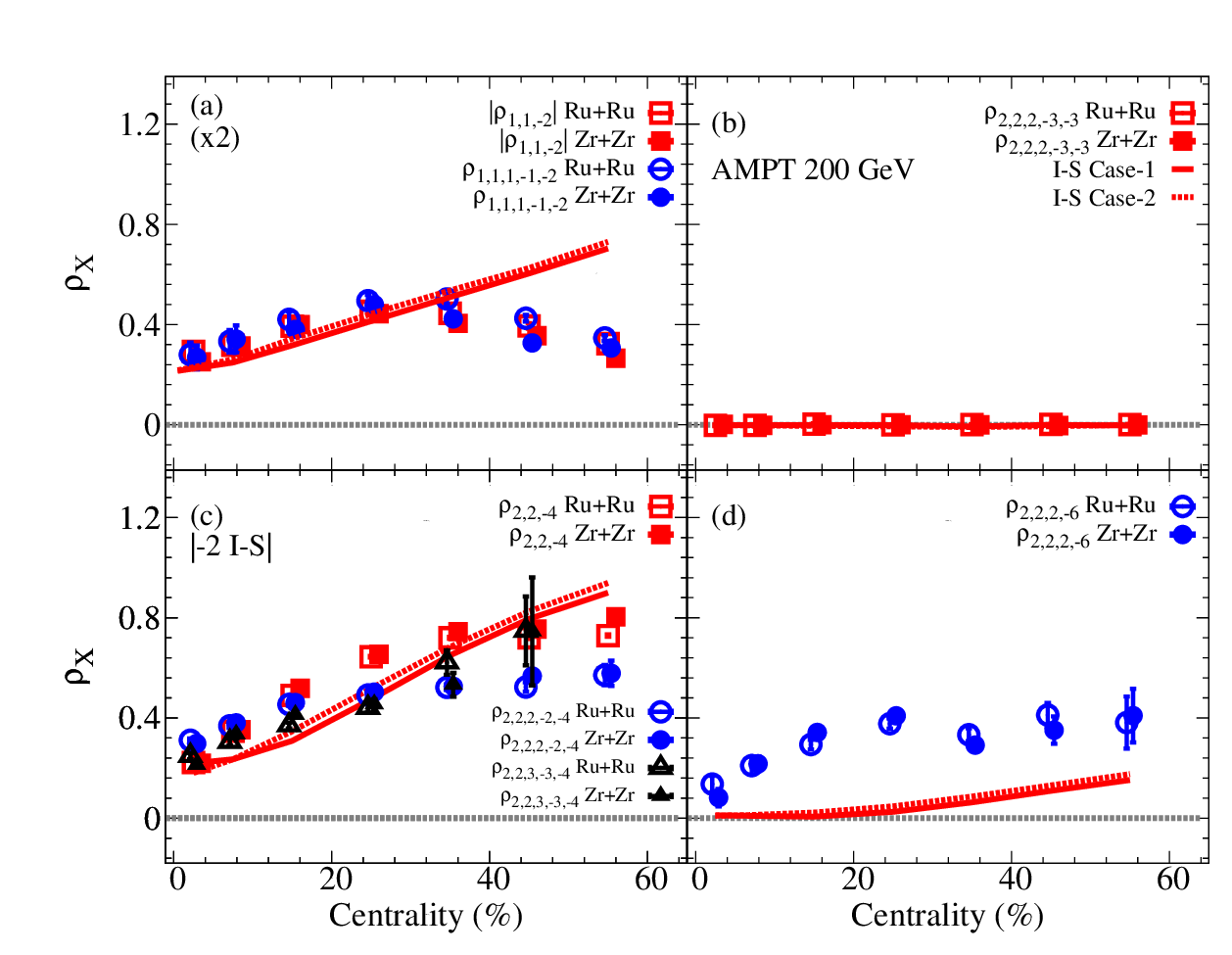}
\vskip -0.4cm
\caption{
Comparisons of the centrality dependence of the normalized asymmetric correlations $\rho_{1,1,-2}$ and $\rho_{1,1,1,-1,2}$ panel (a), $\rho_{2,2,2,-3,-3}$ panel (b), $\rho_{2,2,-4}$, $\rho_{2,2,2,-2,-4}$, and $\rho_{2,2,3,-3,-4}$ panel (c), and $\rho_{2,4,-6}$ panel (d), for \RuRu and \ZrZr at $\sqrt{\textit{s}_{NN}}~=$ 200~GeV from the AMPT model Case-1 and Case-2 parameters given in Tab.~\ref{tab:1}. The curves represent the initial state eccentricity fluctuations.
}\label{fig:r3}
\vskip -0.3cm
\end{figure}
\begin{figure}[!h] 
\includegraphics[width=1.0 \linewidth, angle=-0,keepaspectratio=true,clip=true]{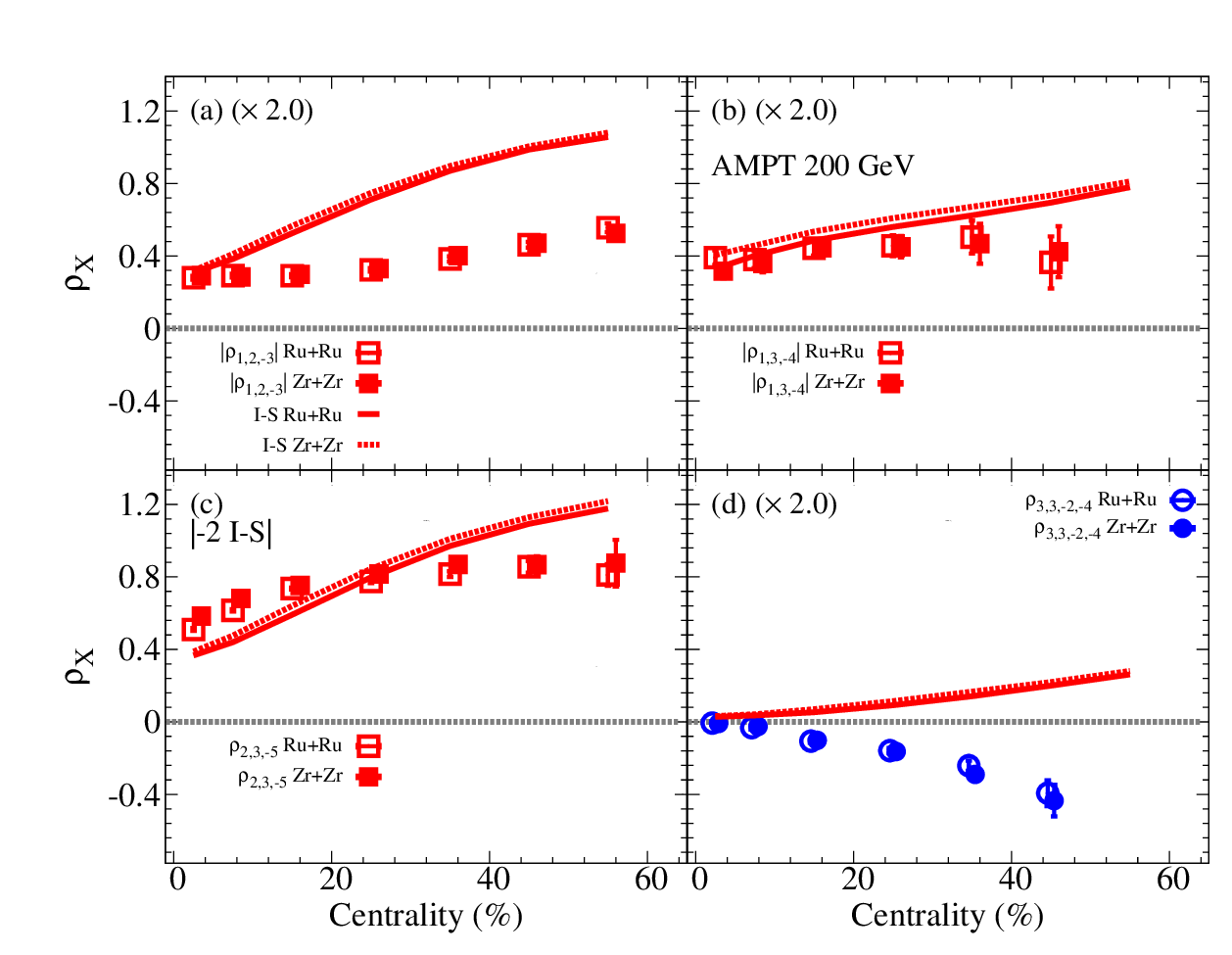}
\vskip -0.4cm
\caption{
Same as in Fig.~\ref{fig:r3} but for $\rho_{1,2,-3}$ panel (a), $\rho_{1,3,-4}$ panel (b), $\rho_{2,3,-5}$, $\rho_{2,3,2,-2,-5}$, and $\rho_{2,3,3,-3,-5}$ panel (c), and $\rho_{3,3,-2,-4}$ and $\rho_{2,3,3,-4,-4}$ panel (d).
}\label{fig:r4}
\vskip -0.3cm
\end{figure}

Figure~\ref{fig:r3} shows the centrality and the system size dependence of the event planes' angular correlations $\langle \cos(2 \psi_{1} - 2 \psi_{2}) \rangle$ (a), $\langle \cos(6 \psi_{2} - 6 \psi_{3}) \rangle$ (b), $\langle \cos(4 \psi_{2} - 4 \psi_{4}) \rangle$ (c) and $\langle \cos(6 \psi_{2} - 6 \psi_{6}) \rangle$ (d) for \RuRu and \ZrZr at $\sqrt{\textit{s}_{NN}}~=$ 200~GeV from the AMPT model Case-1 and Case-2 parameters given in Tab.~\ref{tab:1}.
The ratios $\rho_{1,1,-2}$ and $\rho_{1,1,1,-1,-2}$ are in agreement and increase with centrality selections. The observed agreement between $\rho_{1,1,-2}$ and $\rho_{1,1,1,-1,-2}$ suggest that $\rho_{1,1,-2}$ and $\rho_{1,1,1,-1,-2}$ have a similar contribution from the GMC effects. 
In addition, we found a good agreement in central collisions between the initial and final state constructed $\langle \cos(2 \psi_{1} - 2 \psi_{2}) \rangle$.   
In panel (b), we demonstrated the expected absence of correlations between $\psi_{2}$ and $\psi_{3}$ by presenting the vanishing values of the $\rho_{2,2,2,-3,-3}$ ratios. Our results agree with the expectation that $\psi_{3}$ is a fluctuation-driven event plane. 
The ratios $\rho_{2,2,-4}$, $\rho_{2,2,2,-2,-4}$, and $\rho_{2,2,3,-3,-4}$ panel (c) give the correlation between $\psi_{2}$ and $\psi_{4}$ ($\langle \cos(4\psi_{2} - 4\psi_{4}) \rangle$), the results show a reasonable agreement between the three ratios. 
Also, we presented the positive correlations between $\psi_2$ and $\psi_6$ ($\langle \cos(6\psi_{2} - 6\psi_{6}) \rangle$) given by the ratio $\rho_{2,4,-6}$. 
Our calculations indicated a disagreement in values between the $\rho_{X}$ estimated from initial and final state for $\psi_{2}$--$\psi_{4}$ and $\psi_{2}$--$\psi_{6}$ correlations. 

The system size and centrality dependence of the NASC $\rho_{1,2,-3}$, $\rho_{1,3,-4}$, $\rho_{2,3,-5}$, and  $\rho_{3,3,-2,-4}$ are shown in Fig.~\ref{fig:r4}.
The three event planes correlations $\langle \cos(1\psi_{1} + 2\psi_{2} - 3\psi_{3}) \rangle$, $\langle \cos(1\psi_{1} + 3\psi_{2} - 4\psi_{4}) \rangle$, $\langle \cos(2\psi_{2} + 3\psi_{3} - 5\psi_{5}) \rangle$, and $\langle \cos(6\psi_{3} - 2\psi_{2} - 4\psi_{4}) \rangle$ indicated an increase with centrality selections for both initial and final state correlators. Our results showed a negative correlation for $\langle \cos(6\psi_{3} - 2\psi_{2} - 4\psi_{4}) \rangle$ that disagree with the initial state estimate. Our AMPT calculations for both isobars indicated a disagreement in magnitude between the {\color{black}initial and final state calculations.}

{\color{black}
\subsection{Sensitivity to the deformations and nuclear skin}

Many prior investigations into anisotropic flow observables, especially the ratios involving \RuRu and \ZrZr, have underscored the sensitivity of these observables to nuclear deformations and skin~\cite{Jia:2021tzt, Lu:2023fqd, Jia:2022qgl, Jia:2022qrq, Giacalone:2021uhj, Zhao:2022uhl,Magdy:2023fsp,Magdy:2022cvt,Jia:2022qgl, Lu:2023fqd, Sinha:2023jas, Liu:2022kvz, Li:2019kkh, Xu:2023ges, Xu:2021vpn}. These studies have highlighted that, in central collisions, the lower-order flow harmonics ($n < 4$) exhibit sensitivity to the $\beta_{2}$ and $\beta_{3}$. Higher-order flow harmonics are expected to be sensitive to the interplay between lower- and higher-order deformation parameters. Also, it has been noted that at non-central collisions, the anisotropic flow observables are more susceptible to the change in the nuclear skin.

In this study, our focus is on the absolute values of the observables rather than the \RuRu and \ZrZr ratios. At the absolute values level, our SC results, illustrated in Figs.~\ref{fig:c1}-~\ref{fig:c3}, indicate sensitivity to nuclear deformation in central collisions and nuclear skin in non-central collisions. Figs.~\ref{fig:c4}-~\ref{fig:c11} present the mixed harmonics (A)SC, expected to be sensitive to the interplay between lower- and higher-order deformation parameters in central collisions. Additionally, the NSC and NASC Figs.~\ref{fig:r1}-~\ref{fig:r4}, anticipated to be sensitive to initial-state effects, demonstrate sensitivity to nuclear deformation and skin in central and non-central collisions, respectively. These observations suggest that the presented observables have the potential to constrain the differences in nuclear deformation and skin between $^{96}$Ru and $^{96}$Zr.
}

\section{Summary and outlook}\label{sec:4}
This study utilizes the AMPT model to predict Symmetric Correlations (SC), Asymmetric Correlations (ASC), Normalized Symmetric Correlations (NSC), and Normalized Asymmetric Correlations (NASC) for \RuRu and \ZrZr at RHIC top energy. The presented calculations can be categorized as follows:
\begin{itemize}
    \item {Symmetric and Asymmetric Correlations:}
    The SC and ASC, as depicted in Figs.~\ref{fig:c1}--~\ref{fig:c11}, are anticipated to be sensitive to the interplay between initial and final state effects. Thus, the SC and ASC can validate the assumption that isobaric ratios effectively cancel the final state effect between the two isobars. Moreover, they can constrain the interplay between initial and final state effects in theoretical calculations.
\end{itemize}

\begin{itemize}
    \item {Normalized Symmetric and Asymmetric Correlations:}
    The NSC and NASC, shown in Figs.~\ref{fig:r1}--~\ref{fig:r4}, are predicted to be primarily influenced by initial state effects. Consequently, the NSC and NASC are better suited for studying nuclear structure differences between $^{96}Ru$ and $^{96}Zr$. Our results for \RuRu and \ZrZr reveal a noticeable discrepancy between {\color{black}initial and final state calculations}, emphasizing the necessity for a data-model comparison.
\end{itemize}

In summary, this work presents the centrality dependence of SC, ASC, NSC, and NASC from the AMPT model for \RuRu and \ZrZr at 200 GeV. Our study offers a detailed prediction of these correlations for \RuRu and \ZrZr collisions at RHIC top energy, considering differences in geometry and structure between $^{96}$Ru and $^{96}$Zr. We conclude that conducting detailed comparisons between future experimental measurements and our calculations will aid in constraining the initial and final state effects of the AMPT model.
\section*{Acknowledgments}
The author thanks S. Bhatta and Dr. E. Racow for the valuable discussions and for pointing out essential references.
%
%
This research is supported by the US Department of Energy, Office of Nuclear Physics (DOE NP), under contracts DE-FG02-87ER40331.A008.


\bibliography{ref} 
\end{document}